\documentclass[aps,prl,twocolumn,superscriptaddress,longbibliography,footinbib]{revtex4-1}  %
\usepackage{graphicx} 
\usepackage{float} 
\usepackage{dcolumn} 
\usepackage{bm,color}
\usepackage{amsmath,amssymb,dsfont,amstext,amsfonts}
\usepackage{extarrows}
\usepackage[colorlinks=true,linkcolor=blue,urlcolor=blue,citecolor=blue]{hyperref}
\usepackage{xcolor}
\usepackage{wasysym}
\usepackage{mathtools}
\usepackage{bbold}

\usepackage[normalem]{ulem} 
\usepackage{color}

\newcommand{\la}{\langle}
\newcommand{\ra}{\rangle}

\newcommand{\bk}{\bm{k}}

\def\dd{\mathrm{d}}

\begin{document}
\title{Hopf characterization of two-dimensional Floquet topological insulators}
\author{F. Nur \"{U}nal}
\email{unal@pks.mpg.de}
\affiliation{Max-Planck-Institut f{\"u}r Physik komplexer Systeme, N{\"o}thnitzer Stra\ss e 38, 01187 Dresden, Germany}
\author{Andr\'{e} Eckardt}
\email{eckardt@pks.mpg.de}
\affiliation{Max-Planck-Institut f{\"u}r Physik komplexer Systeme, N{\"o}thnitzer Stra\ss e 38, 01187 Dresden, Germany}
\author{Robert-Jan Slager}
\email{robertjanslager@fas.harvard.edu}
\affiliation{Department of Physics, Harvard University, Cambridge, MA 02138}

\date{\today}

\begin{abstract}
We present a topological characterization of time-periodically driven two-band models in $2+1$ dimensions as Hopf insulators. The intrinsic periodicity of the Floquet system with respect to both time and the underlying two-dimensional momentum space constitutes a map from a three dimensional torus to the Bloch sphere. As a result, we find that the driven system can be understood by appealing to a Hopf map that is directly constructed from the micromotion of the drive. Previously found winding numbers are shown to correspond to Hopf invariants, which are associated with linking numbers describing the topology of knots in three dimensions. Moreover, after being cast as a Hopf insulator, not only the Chern numbers, but also the winding numbers of the Floquet topological insulator become accessible in experiments as linking numbers. We exploit this description to propose a feasible scheme for measuring the complete set of their Floquet topological invariants in optical lattices.
\end{abstract}

\maketitle

\paragraph{Introduction --}
With the advent of topological insulators \cite{HasanKane10_RMP, Qi11_RMP}, the past decades have witnessed a rekindling of interest in band theory. The interplay between symmetry and topology has led to the prediction and observation of many novel gapped and semi-metallic electronic topological phases. While the topological classification of such free-fermion systems is by now rather well understood  \cite{Kitaev09_AIP, Ryu10_NJP, Fu11_PRL, Slager12_NatPhys, Kruthoff17_PRX, Po17_NatCommun, Holler18_PRB, Bouhon2018_arxiv,Bradlyn17_Nat,Slager19_JPCS,Juricic12_PRL_Slager}, notions and invariants such as Chern numbers have  increasingly found generalizations  in the context of periodically driven quantum systems~\cite{Roy17_PRB,Kitagawa10_PRB,Rudner13_PRX,Unal18_arx}. A striking result was found recently also for quenched systems~\cite{WangZhai17_PRL,Tarnowski17_arx,SunPan18_PRL,Zhang18_SciBulletin,Yu17_PRA,Yuan17_ChPL}. Particularly, by using a composition map amounting to a Hopf map, it was shown that the Chern number can be directly understood as a linking number.
Such out-of-equilibrium topological characterizations are not only intriguing from a theoretical perspective, but are also increasingly finding their way to experimental settings of ultracold atoms via Floquet engineering, where the Berry curvature and Chern numbers
of Bloch bands
have been measured~\cite{Schneider16_Sci,Aidelsburger15_NatPhys,Flaschner18_Nat,Flaschner16_Sci,Jotzu14_Nat, Tarnowski17_arx,Aidelsburger13_PRL,Miyake13_PRL,Eckardt17_RMP}.

Nonetheless, while the topological characterization of non-equilibrium periodically driven lattices in two dimensions was established theoretically, the winding numbers characterizing these systems have not been observed directly in experiments so far, whereas the associated anomalous edge states were probed in photonic systems~\cite{HuPillay15_PRX,GaoGao16_NatCommun,Mukherjee17_NatComm,Mukherjee18_NatComm,Maczewsky17_NatCommun}.
Here we reveal that the full topological characterization of the system can be achieved via a simple but universal scheme involving Hopf maps. Specifically, the winding numbers are shown to directly correspond to Hopf invariants. In contrast to the symmetry protected topological phases, topology of Hopf insulators relate to the topology of knots formed in three dimensions under the Hopf map. That is, whereas symmetry-protected topological band theory involves K-theory and hence is stable to the addition of higher bands, Hopf insulators coincide with a Hopf map by virtue of being a two-band model. Specifically, the Grassmannian target manifold $Gr(m,m+n)=Gr(1,2)$ is topologically the same as a sphere $S^2$. Several proposals involving three-dimensional crystals~\cite{LiuVafa17_PRB,Kennedy16_PRB,Deng13_PRB,Moore08_PRL,HeChih-Chun19_PRB} and dipolar gases~\cite{Schuster19_Hopfdipolar} have been theoretically studied and proposed as platforms to identify this novel topologically insulating state, however, with no luck in an experimental observation so far. In this paper, we consider a two-band model in two dimensions under a Floquet drive which provides for the third periodicity necessary for the Hopf map. We show that the Hopf invariant of the micromotion fully captures the non-trivial winding structure of the Floquet system. We find that upon varying the period of the drive, in combination with a flattening procedure, the Hopf characterization can be experimentally invoked to deduce the full set of Floquet topological invariants of the two-dimensional quantum system. We thus propose a feasible approach for measuring the winding numbers as linkings in the dynamical response of the system without requiring adiabatic ground-state preparation.

\begin{figure}
	\centering\includegraphics[width=1\linewidth]{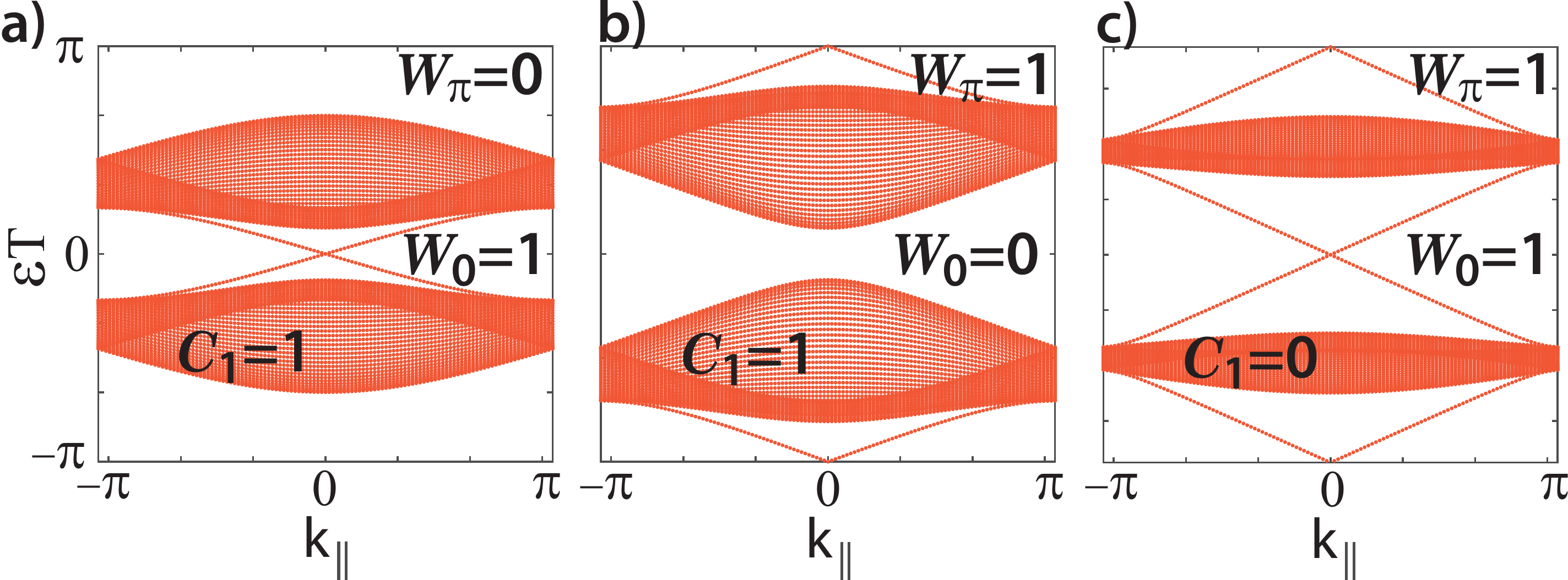}
	\caption{ Quasienergy spectrum of a step-wise drive introduced in Eq.(\ref{eq-Hamiltonian}) in a strip geometry for momentum $k_{\parallel}$ along the periodic direction. Three different combinations of edge states appear in the $0$ and/or the $\pi$-gap for a) $T=2\pi/3J,\: \Delta=0$; b) $T=2\pi/2J,\: \Delta=1.3J$; c) $T=2\pi/1.5J,\: \Delta=0.1J$.
	 }
	\label{fig_cases}
\end{figure}

\paragraph{Model setting --}
We start from a generic two-band model in two dimensions,
\begin{equation}
{\cal H}(\bk,t)=\bm{h}(\bk,t)\cdot\bm{\sigma},
\end{equation}
with the Pauli matrices $\bm{\sigma}$ and quasi-momenta $\bk$. Under a periodic drive with period $T$, the time-evolution of the system is captured by the time-ordered operator $\mathcal{U}(\bk,t)=\mathcal{T}\exp[-i\int_{0}^t{\cal H}(\bk,t')\dd t']$. Evaluated stroboscopically, $\mathcal{U}(\bk,T)=e^{-i \mathcal{H}_F(\bk) T}$, it defines the quasienergy spectrum of the system through the time-independent Floquet Hamiltonian $\mathcal{H}_F$ with eigenvalues $\varepsilon_n$ for the two bands $n=1,2$. Due to the periodic nature of the Floquet spectrum, quasienergies can only be defined modulo $2\pi$ and can be restricted to the Floquet Brillouin Zone (FBZ), $-\pi/T<\varepsilon_n\leq\pi/T$. We label the two gaps centered around the quasienergy $g/T$ by $g=0$ and $g=\pi$. With the quasienergy spectrum defined on a circle, it is possible to obtain anomalous edge states lying at the $\pi$-gap connecting the bands through the FBZ edge. This renders the equilibrium topological classification in terms of Chern number $C_n$ inept to characterize driven systems (see Fig.\ref{fig_cases}). Instead, one needs to consider winding numbers, $W_g$, which are topological invariants associated with gaps rather than individual bands~\cite{Rudner13_PRX,Kitagawa10_PRB}.

The time evolution operator can be decomposed into two parts,
\begin{equation}\label{eq-micromotion}
\mathcal{U}(\bk,t)=\mathcal{U}_F(\bk,t)e^{-i\mathcal{H}_F(\bk)t},
\end{equation}
corresponding to the effective Floquet Hamiltonian $\mathcal{H}_F$ dictating the stroboscopic evolution at the end of a period $T$, and the time-periodic micromotion operator $\mathcal{U}_F$ capturing the details within a period, whose stroboscopic effect equals the identity $\mathcal{U}_F(T)=\mathds{1}$. In the high-frequency regime where the compactness of the FBZ becomes irrelevant, the micromotion describes only trivial corrections and we can focus on the evolution captured by the Floquet Hamiltonian. However, the overall effect of the micromotion can be non-trivial at lower frequencies by winding through the FBZ~\cite{Rudner13_PRX}, giving rise to non-zero winding number $W_{\pi}$.

For concreteness, we consider a two-dimensional honeycomb lattice with an energy offset $\Delta$ between the two sublattices, and focus on a step-wise periodic drive~\cite{Kitagawa10_PRB,Quelle17_NJP}. Driving is introduced by dividing one period into three stages of equal length $T/3$ during which the tunneling parameters are switched on and off in a cyclic manner. The momentum-space tight-binding Hamiltonian can be written as
\begin{equation}\label{eq-Hamiltonian}
{\cal H}(\mathbf{k},t)=-\sum_{i=1}^3 J_i(t) \big\{\cos(\mathbf{k}\cdot\mathbf{b}_i)\sigma_x +\sin(\mathbf{k}\cdot\mathbf{b}_i)\sigma_y \big\} +\frac{\Delta}{2}\sigma_z,
\end{equation}
where the nearest-neighbor tunneling amplitudes $J_i$ are along the directions $\mathbf{b}_1=(-1,0),\: \mathbf{b}_2=(\sqrt{3}/2,1/2)$ and $\mathbf{b}_3=(-\sqrt{3}/2,1/2)$ in units of the nearest-neighbor distance. During the $i^{th}$ stage within a period, only the hopping along the $b_i$ direction is allowed with an amplitude $J_i=J$ while the hopping amplitudes along the other two directions are set to zero.
We note, however, that we obtain equivalent results for a (continuous) circularly-driven honeycomb lattice. The former has the advantage of theoretical simplicity whereas the latter has been already implemented in cold-atom experiments~\cite{Flaschner18_Nat,Tarnowski17_arx,Jotzu14_Nat}. Even though such stepwise drives were initially introduced to demonstrate the topological distinction of nonequilibrium periodically driven systems, they have been already realized in photonic wave guides, and their implementation in cold atoms is also possible~\cite{Quelle17_NJP}.

The cyclic nature of the allowed tunneling directions breaks time-reversal symmetry and may induce chiral edge states in the quasienergy gaps $g$. Fig.~\ref{fig_cases} demonstrates the three possible combinations of edge states for an armchair-terminated strip with momentum $k_{\parallel}$ along the periodic direction. Particularly, as can be seen in Fig.~\ref{fig_cases}(c), the anomalous edge states appearing at the FBZ edge result in the break-down of the conventional bulk-boundary correspondence of static systems \cite{Slager15_Bbc,Hatsugai93_Bbc,Slager16_gbbc,Essin11_Bbc,Rhim18_Bbc}. Instead, the bulk-boundary correspondence of Floquet systems is captured by the winding number $W_g$ that accounts for the winding structure of the micromotion in contrast to the Chern number that can be associated to the effective Floquet Hamiltonian.

We now turn to the topological aspects alluded to above.
We consider the evolution of a topologically trivial state $\Psi(\bk,t=0)$ under the effect of a time-periodic Hamiltonian (\ref{eq-Hamiltonian}) throughout an entire period $0<t\leq T$. In two spatial dimensions, the quasimomentum $\bk$ defines a two-dimensional torus $T^2$. Under the micromotion $\mathcal{U}_F$ of the drive, the state returns to itself at the end of a period, $\Psi(\bk,T)=\mathcal{U}_F(\bk,T)\Psi(\bk,0)=\Psi(\bk,0)$, making the evolution periodic and defining a three-torus $T^3$ in $\bm{p}\equiv(k_x,k_y,t)$-space. It is this periodicity in all three parameters that allows for establishing the topological characterization of the periodically driven system in terms of a Hopf map, which pertains to a map from $S^3$ to $S^2$ and arises by virtue of $\pi_3(S^2)=\mathbf{Z}$~\cite{Pontryagin41_hopf,Makhlin95_Jetp}. Mathematically, it can be generalized to $T^3$ by use of a composition map from $T^3$ to $S^3$. This restrains the map to have values in a finite group $\mathbf{Z}_{2GCD(C_x,C_y,C_t)}$ in terms of the greatest common divisor of the Chern numbers $C_i$ of the respective two-torus submanifolds within the three-dimensional torus formed by $\bm{p}$. Assuming these `weak invariants' to be absent then renders the usual $\mathbf{Z}$-valued characterization. Specifically, the micromotion defines an evolution on the Bloch sphere $S^2$, whose topology is described by the Hopf map conventionally given as
\begin{equation*}
H=-\frac{1}{16\pi^2}\int A\wedge B,
\end{equation*}
where $A_i=-2i\Psi^{\dag} \partial_i\Psi$ is the connection one-form and $B=dA=\frac{1}{2}B_{ij}dp^i\wedge dp^j$, with $B_{ij}=-2i(\partial_i\Psi^{\dag} \partial_j\Psi-\partial_j\Psi^{\dagger}\partial_i\Psi)$, entails the Hopf curvature two-form. We can cast this Hopf invariant into a more practical form~\cite{Ren07_JMathPhys}
\begin{equation}\label{eq-Hopf}
H=-\frac{1}{4\pi^2}\int \dd^3p\; \epsilon^{ijk}\; \Psi^{\dag} \partial_i\Psi \partial_j\Psi^{\dag} \partial_k\Psi,
\end{equation}
which we shall use in the following.

\begin{figure}
	\centering\includegraphics[width=1\linewidth]{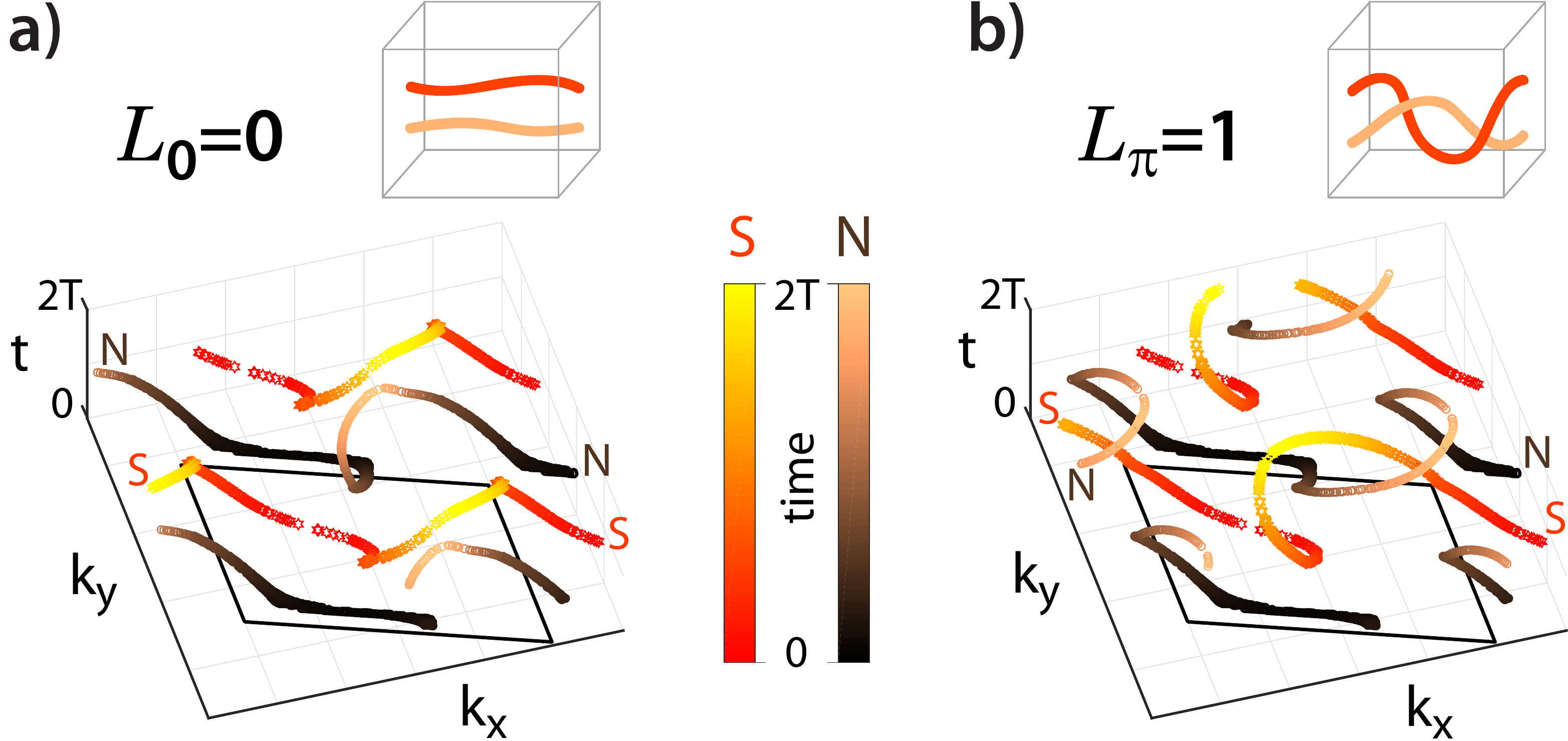}
	\caption{ Numerical calculation of the linking number $L_g$ associated to each quasienergy gap of Fig.~\ref{fig_cases}(b). A state evolving under the periodic drive (\ref{eq-returnMap}) visits the north (N) and the south (S) poles of the Bloch sphere at $\bk$-points marked by different colors and times encoded by gradient. These inverse images of the poles draw trajectories which (a) do not link at the central gap $g=0$ and (b) link once within the momentum-space BZ marked by the parallellogram at the $g=\pi$-gap. The insets depict the linking in $T^3$ schematically.}
	\label{fig_linking}
\end{figure}

Under the Hopf construction, each point $\bm{p}$ maps to a vector on $S^2$, where the inverse images of any two vectors elucidate the topological invariant as a linking number $L$ in $T^3$ (see insets in Fig.~\ref{fig_linking}). That is, the pre-images define linking circles and lines in the case of a non-trivial Hopf characterization. This correspondence has been previously employed to measure the Chern number of the Floquet bands in stroboscopic measurements~\cite{Tarnowski17_arx}. Here, we show that the linking number of the micromotion provides direct access to the winding number of Floquet topological insulators and, hence, captures the full topological classification of the periodically driven system.

We emphasize the crucial role of the periodicity in the system. The three-dimensional torus is formed by the micromotion, under the evolution of which the state has to be mapped onto itself at the end of a period. That is, the micromotion has to be isolated from the stroboscopic evolution by smoothly deforming the drive to obtain $\mathcal{U}(T)=\mathcal{U}_F(T)=\mathds{1}$, i.e.\ degenerate quasienergy bands. This can be achieved by multiplying Eq.(\ref{eq-micromotion}) with $e^{i\mathcal{H}_FT}$ from the right. However, there is an ambiguity in determining the effective Floquet Hamiltonian $\mathcal{H}^g_F(\bk)=i\log{\mathcal{U}(\bk,T)}/T$ when choosing different branches of the logarithm lying at gap $g$. The micromotion operator associated with a chosen Hamiltonian can be implemented by evolving the system with $-\mathcal{H}^g_F$ for an amount of time $T$ after completing one period of the drive~\cite{Rudner13_PRX},
\begin{equation}\label{eq-returnMap}
\mathcal{U}^g_F(\bk,t)=
\begin{cases}
\mathcal{U}(\bk,t),    & 0< t\leq T,\\
e^{-i\mathcal{H}^g_F(\bk)(T-t)},      & T< t\leq 2T,
\end{cases}
\end{equation}
with a rescaling of the period. Physically, Eq.(\ref{eq-returnMap}) corresponds to contracting the bands ($\mathcal{U}(T)=\mathds{1}$) into degeneracy at the center or at the edge of the FBZ. When we apply $\mathcal{U}^g_F$ to a topologically trivial initial state, this state acquires the periodicity of the micromotion, thereby naturally tracing an evolution of the well-defined Bloch vector as function of the variables $\bm{p}$ on $T^3$, governed by the Hopf map. Consequently, the driven system is described by two Hopf constructions, hence, by two Hopf invariants $H_g$ depending on whether the branchcut is taken along $\pi$ or $0$. The former reveals the topology of the Hopf map at the $\pi$-gap as the micromotion winds the state through the zone edge, 
whereas the latter corresponds to the Hopf invariant at the zero gap
.

As one of our main results, we find that our viewpoint naturally encloses the previously found topological characterizations by connecting to the winding number pertaining to the gap selected by the branchcut. Indeed, by writing the evolution operators in the basis of Bloch vectors, we analytically calculate that the winding number~\cite{Rudner13_PRX,Kitagawa10_PRB},
\begin{equation}\label{eq-Winding}
W_g=
\int \frac{d^3p}{24\pi^2}
\:\varepsilon^{ijk}\:\mathrm{Tr}[(\mathcal{U}^{g\dag}_F\partial_{i} \mathcal{U}^g_F) (\mathcal{U}^{g\dag}_F\partial_{j}\mathcal{U}^g_F) (\mathcal{U}^{g\dag}_F\partial_{k}\mathcal{U}^g_F)],
\end{equation}
directly coincides with the Hopf characterization\footnote{See Supplementary for the proof.}, 
\begin{equation}
H_g=W_g.
\end{equation}
Hence, we not only provide an experimentally viable realization for a Hopf insulator, but also establish a general perspective on its bulk-boundary correspondence via the winding numbers. Moreover, the difference between these two Hopf invariants gives the Chern number of the quasienergy bands.

To make these statements concrete, we can revert to the explicit values of these topological invariants in the setting of our illustrative model. We numerically calculate the Hopf invariants (winding numbers) given in Fig.~\ref{fig_cases}, which can be directly observed as linking numbers, $H_g=L_g$, of the inverse images of the north and south poles of the Bloch sphere in the three-dimensional $\bm{p}$-space. In Fig.~\ref{fig_linking}, we follow the evolution of an initial state defined by the Bloch vector $\Psi(\bk,0)=(1,3,0)/\sqrt{10}$ after suddenly switching on the drive with the parameters given in Fig.~\ref{fig_cases}(b), and plot the $\bm{p}$-values where the state $\Psi(\bk,t)$ points along the poles. Upon employing the return map, when we choose the branchcut for which the quasienergy bands are degenerate at the FBZ edge, the trajectories in Fig.~\ref{fig_linking}(a) do not link, corresponding to the vanishing winding number at the $0$-gap, $L_0=W_0=0$. However, when the quasienergy bands are contracted into degeneracy at the FBZ center, the quasienergy spectrum features edge states winding through the zone edge. Correspondingly, the trajectories in Fig.~\ref{fig_linking}(b) link once, giving $L_{\pi}=W_{\pi}=1$.

\paragraph{Experimental Scheme --}
We now turn to the experimental implications of our characterization.
The trajectories depicted in Fig.~\ref{fig_linking} can be measured experimentally in optical lattices via the state-tomography technique~\cite{Hauke14_PRL,Flaschner18_Nat} where the inverse images of the poles appear as vortices in the azimuthal Bloch-sphere angle of the state $\Psi(\bk,t)$ as a function of $\bk$. This method has been recently employed in a circularly driven honeycomb lattice to measure the Chern number with great precision~\cite{Tarnowski17_arx}, by monitoring the evolution of the singularities stroboscopically. However, since the Chern numbers of the two bands, are given by the difference of the winding numbers above and below a band,
\begin{equation}\label{eq-Chern-Winding}
C_n=(-1)^n(W_{\pi}-W_0),
\end{equation}
this measurement is not enough to reveal the full topological characterization of a Floquet system. Here we propose a different scheme that relies on tracing the evolution within one period $0<t\leq T$.
Measurement of the linking number requires the isolation of the micromotion from the stroboscopic dynamics to be able to observe $\mathcal{U}^g_F(T)=\mathds{1}$. In principle, this can be achieved by designing the return map depicted in Eq.~(\ref{eq-returnMap}). However in practice, it is not possible to engineer $\mathcal{H}^g_F$ as static Hamiltonian. Alternatively, here we propose a more elegant and feasible approach to extract the micromotion in the experiments directly.

\begin{figure}
	\centering\includegraphics[width=1\linewidth]{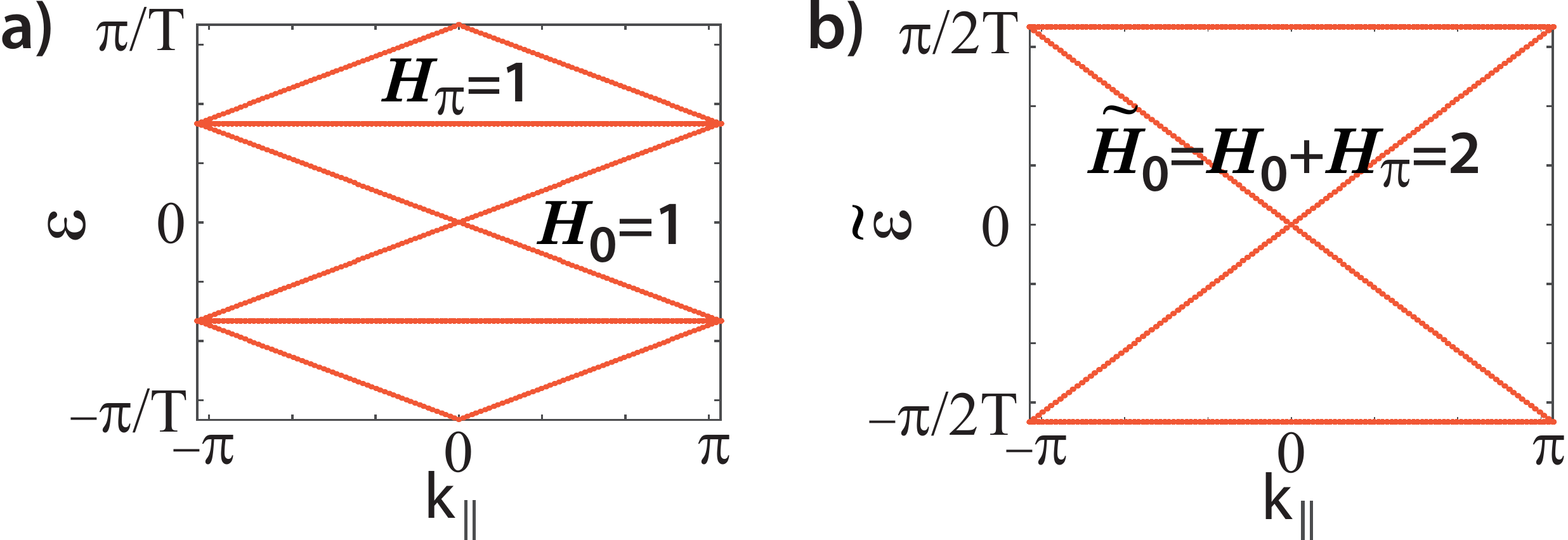}
	\caption{ Folding of the quasienergy spectrum in the strip geometry when the period is doubled, for $JT/3=\pi/2,\,\Delta=0$. The bulk Hopf invariant of the central gap over the double period $\tilde{T}=2T$ is the sum of the topological invariants of the both gaps of the original drive, $\tilde{H}_0=H_0+H_{\pi}$, which can be directly revealed as the linking number.}
	\label{fig_doublePeriod}
\end{figure}

Our scheme relies on the particle-hole symmetry of the Hamiltonian and on folding the quasienergy spectrum, bringing the clear advantage of not relying on adiabatic preparation of the topological ground state (cf.~\cite{Unal18_arx}). When the drive is monitored in terms of a doubled period $\tilde{T}=2T$ (with respect to which it is, obviously, also periodic), the quasienergy bands fold once around the energy $\varepsilon=\pm \pi/2T$. As a result, the edge states at the $\pi$-gap of the original drive lie at the zero-gap of the new quasienergy spectrum of period $2T$. This also implies that the winding number in the zero-gap of the folded Floquet spectrum is given by the sum of the two winding numbers of the original drive of period $T$; i.e.\ $\tilde{W}_0^{(2T)}=W_0^{(T)}+W_{\pi}^{(T)}$. Taking power from this observation, we now focus on making the quasienergy bands flat at $T\varepsilon=\pm \pi/2$, so that they become degenerate when the period is doubled as demonstrated in Fig.~\ref{fig_doublePeriod}.

In principle, the quasienergy bands can be adiabatically flattened by slightly deforming any drive without closing any gaps as for a static system; e.g.~Fig.~\ref{fig_doublePeriod}(a) is topologically identical to the Floquet spectrum given Fig.~\ref{fig_cases}(c).
This can be achieved by tuning the driving frequency $\omega$ \cite{Unal18_arx} individually at each quasimomentum $\bk$ to shift the quasienergies within the FBZ~\footnote{One can imagine this by comparing with the high-frequency limit, where the bands are concentrated in a region ${J}\ll\omega$ around zero. As the driving frequency is lowered, the two energy scales become comparable ${J}\approx\omega$ and the bands expand within the FBZ. See Supplementary for details.}. We identify this modified drive with a tilde, e.g.~$\tilde{{\cal H}}(\bk,t)$, as well as its topological invariants. In practice, the deformation needs to be performed only at a given number of points in the BZ enough to reveal the trajectories of the vortices. Moreover, this experimental implementation does not require the knowledge of the theoretical model, since the quasienergy gap can be found experimentally from the stroboscopic evolution after quenches~\cite{Unal18_arx}.

Once the quasienergies are flattened, the time evolution operator over a double period becomes the identity, $\tilde{\mathcal{U}}(2T)=\mathds{1}$, directly corresponding to the micromotion operator $\tilde{\mathcal{U}}^0_F(2T)$. Fig.~\ref{fig_doublePeriod}(b) demonstrates the degenerate Floquet spectrum over the doubled period where the linking number corresponds to the winding number at the zero-gap, $\tilde{L}^{(2T)}=\tilde{W}_0^{(2T)}$.
This linking number can be measured via state-tomography by tracing the time evolution of the state (i.e.~vortices in momentum-space azimuthal-phase profile) throughout $0<t\leq 2T$. In terms of the winding numbers of the original Hamiltonian, this corresponds to their sum,
\begin{equation}\label{eq-linkingDouble}
\tilde{L}^{(2T)}=W_0^{(T)}+W_{\pi}^{(T)},
\end{equation}
which is, on its own, still not enough to identify the winding numbers individually. The missing information can be acquired by combining the equations (\ref{eq-linkingDouble}) for the micromotion and the Chern number (\ref{eq-Chern-Winding}) which can be obtained via a stroboscopic measurement~\cite{Tarnowski17_arx}.

\paragraph{Conclusion and Discussion --}
We have shown that two-dimensional Floquet topological insulators with two bands are characterized by two Hopf constructions rooted in the inherent periodicity of micromotion. The associated Hopf invariants are found to directly correspond to previously described winding numbers and, thus, provide a complete topological characterization of the system. The attained perspective is not only appealing from a theoretical point of view, employing Hopf invariants to describe the quasienergy band topology, but also intimately relates to experimental setups, bringing these deep notions within experimental reach. Indeed, by proposing
a viable band folding scheme, this physics is directly measurable in cold-atomic systems.

Finally, we point out a peculiarity compared to static systems \cite{Deng13_PRB, LiuVafa17_PRB, Moore08_PRL}. The three dimensional $\mathbf{Z}$-valued Hopf invariant  is not present in the many-band tenfold periodic table
In contrast, when a two-dimensional system is driven, the third periodicity arises by virtue of time translations. The periodic table for Floquet states \cite{Roy17_PRB}, predicts a $\mathbf{Z}^{\times n}$ classification, where $n$ refers to the number of gaps, showing e.g. that there is a $0$- and $\pi$-gap topological invariant for the case of two bands. As we have demonstrated, these numbers directly coincide with our framework and the Hopf characterization of the micromotion, highlighting the difference between $(2+1)$-dimensional Floquet states and $3$ dimensional static insulators, as reflected in the periodic tables themselves. This gives incentive to speculate that our approach can be generalized to describe the topology of any desired pair of bands in a general two-dimensional class A Floquet insulator. In this regard we remark that in specific scenarios the Hopf maps can be obtained by considering stacks of Chern insulators or can be stabilized to higher bands under certain symmetries that can be conserved in the protocol at hand~\cite{Pontryagin41_hopf,Makhlin95_Jetp,Kennedy16_PRB, LiuVafa17_PRB}. This poses the question whether the Chern number of quasienergy bands can be phrased into such a ``stacking" perspective in the time direction, using e.g.~a Hopf map that is the composition of the Hopf maps for both gaps. Indeed, from our construction it directly follows that restrictions on the Hopf invariant correspond to Chern numbers of cuts and vice versa. Finally, this also stimulates connections to Topological Hopf-Chern insulators \cite{Kennedy16_PRB}, which will in turn relate to the many-band generalization.

\begin{acknowledgments}
	{\it Acknowledgments --}
	F.N.\"{U} and A.E.~acknowledge the support from the Deutsche Forschungsgemeinschaft (DFG) via the Research Unit FOR 2414 (under Project No.~277974659) and fruitful discussions with Christof Weitenberg. R.-J.S gratefully acknowledges funding via Ashvin Vishwanath from the Center for the Advancement of Topological Materials initiative, an Energy Frontier Research Center funded by the U.S. Department of Energy, Office of Science.
\end{acknowledgments}

{\it Note added --}
 While finalizing this work, we became aware of the insightful work by Schuster et al. \cite{Schuster19_FHopf} which, in contrast to the present paper, considers the Floquet driving of an underlying three-dimensional Hopf insulator. Such systems were shown to have an additional $\mathbf{Z}_2$ invariant associated with the Witten anomaly\cite{HeChih-Chun19_PRB}.

\bibliography{references}


\pagebreak

\vspace{3cm}
\onecolumngrid
\begin{center}
  \section{ Supplementary for ``Hopf characterization of two-dimensional Floquet topological insulators"}
\end{center}

\setcounter{equation}{0}
\setcounter{figure}{0}
\setcounter{table}{0}
\setcounter{page}{1}
\makeatletter
\renewcommand{\theequation}{S\arabic{equation}}
\renewcommand{\thefigure}{S\arabic{figure}}

\subsection{A. Hopf Invariant and Winding Number Correspondence}
In this supplementary, we analytically prove that the Hopf invariant given in Eq.~(4) of the main text is equal to the winding number previously studied in the literature. For this purpose, we start by contracting the winding number given in Eq.~(6) into the form
\begin{equation}\label{eq-Winding}
W= -\frac{1}{24\pi^2}
\int d^3p\; \varepsilon^{ijk}\mathrm{Tr}[U^\dag \partial_{i} U\cdot \partial_{j}U^\dag \cdot \partial_{k}U],
\end{equation}
by using the unitarity of the evolution operator, $UU^\dag=1$ and $\partial_{i}U\cdot U^\dag+ U\partial_{ji}U^\dag=0$. Following the Einstein notation, the trace can be written as
\begin{equation}
W= -\frac{1}{24\pi^2}
\int d^3p\; \varepsilon^{ijk}
\sum_{a,b,c,d} U^\dag_{ab} \partial_{i} U^{\phantom{\dag}}_{bc} \partial_{j}U^\dag_{cd} \partial_{k}U^{\phantom{\dag}}_{da},
\end{equation}
where the indices $a,b,c,d=1,2$ scan the space formed by the two bands. Matrix elements of the evolution operator can be found by evolving an initial state $|\psi_a\rangle$ which we take to be $\bk$-independent, $U_{ab}(\bk,t)=\langle\psi_a|U(\bk,t)|\psi_b\rangle$. The derivatives act only on the time-evolved state which we denote as $|\Psi_a(\bk,t)\rangle=U(\bk,t)|\psi_b\rangle$,
\begin{align}
W&= -\frac{1}{24\pi^2}
\int d^3p\; \varepsilon^{ijk}
\sum_{a,b,c,d} \la\Psi_a(\bk,t)|\psi_b\ra \partial_{i} \la\psi_b|\Psi_c(\bk,t)\ra \partial_{j}\la\Psi_c(\bk,t)|\psi_d\ra  \partial_{k}\la\psi_d|\Psi_a(\bk,t)\ra, \nonumber\\
&= -\frac{1}{24\pi^2}
\int d^3p\; \varepsilon^{ijk}
\sum_{a,c} \la\Psi_a(\bk,t)|  \partial_{i}\Psi_c(\bk,t)\ra  \la\partial_{j}\Psi_c(\bk,t)| \partial_{k}\Psi_a(\bk,t)\ra.
\end{align}
Here, we dropped the terms equal to identity $\sum_a |\psi_a\ra\la\psi_a|=\mathds{1}$. Therefore, we identify the four terms giving contribution to the winding numbers as,
\begin{equation} \label{eq-Winding-4terms}
W= -\frac{1}{24\pi^2}
\int d^3p \varepsilon^{ijk} \Big[
 \underbrace{\Psi_1^\dag \partial_{i}\Psi_1 \partial_{j}\Psi_1^\dag \partial_{k}\Psi_1}_{W_1} +
 \underbrace{\Psi_1^\dag \partial_{i}\Psi_2 \partial_{j}\Psi_2^\dag \partial_{k}\Psi_1}_{W_2} +
 \underbrace{\Psi_2^\dag \partial_{i}\Psi_1 \partial_{j}\Psi_1^\dag \partial_{k}\Psi_2}_{W_3} +
 \underbrace{\Psi_2^\dag \partial_{i}\Psi_2 \partial_{j}\Psi_2^\dag \partial_{k}\Psi_2}_{W_4} \Big].
\end{equation}
One can intuitively see that $W_1=W_4$ and $W_2=W_3$ by using symmetry arguments but we also prove it below. The important point is to relate the mixed term $W_2$ to the first term $W_1$ which is already in the same form as the Hopf invariant Eq.(4).

The time-evolved wave functions $\Psi_a(\bk,t)$ are two-component normalized spinors on the Bloch sphere. They can be written in terms of two complex scalar fields $z_1(\bk,t)$ and $z_2(\bk,t)$ as $\Psi_1(\bk,t)=(z_1(\bk,t),z_2(\bk,t))^T$ and $\Psi_2(\bk,t)=(-z_2(\bk,t),z_1(\bk,t))^T$, satisfying the normalization condition $|z_1(\bk,t)|^2+|z_2(\bk,t)|^2=1$ at each $\bk$ and $t$. We insert these forms into the first term in Eq.~(\ref{eq-Winding-4terms}),
\begin{align} \label{eq-W1}
W_1&= (z_1,  z_2)(\partial_{i}z_1, \partial_{i}z_2)^T  (\partial_{j}z_1, \partial_{j}z_2) (\partial_{k}z_1, \partial_{k}z_2)^T \nonumber\\
&= z_1^*\partial_iz_1\partial_jz_1^*\partial_kz_1 + z_1^*\partial_iz_1\partial_jz_2^*\partial_kz_2 +
  z_2^*\partial_iz_2\partial_jz_1^*\partial_kz_1 + z_2^*\partial_iz_2\partial_jz_2^*\partial_kz_2 \nonumber\\
&= z_1^*\partial_iz_1\partial_jz_2^*\partial_kz_2 + z_2^*\partial_iz_2\partial_jz_1^*\partial_kz_1,
\end{align}
where the terms symmetric in $ijk$ vanish because of the levi-civita operator in the integral, e.g.\ $z_1^*\partial_iz_1\partial_jz_1^*\partial_kz_1=-z_1^*\partial_kz_1\partial_jz_1^*\partial_iz_1=0$. It can be directly seen that $W_1=W_4$ when we write the $W_4$ term like
\begin{equation} \label{eq-W4}
W_4= z_2^*\partial_iz_2\partial_jz_2^*\partial_kz_2 + z_2^*\partial_iz_2\partial_jz_1^*\partial_kz_1 +
  z_1^*\partial_iz_1\partial_jz_2^*\partial_kz_2 + z_1^*\partial_iz_1\partial_jz_1^*\partial_kz_1 = W_1.
\end{equation}

Similarly,
\begin{align} \label{eq-W2}
W_2&= z_1^*\partial_iz_2^*\partial_jz_2\partial_kz_1 - z_1^*\partial_iz_2^*\partial_jz_1\partial_kz_2 -
  z_2^*\partial_iz_1^*\partial_jz_2\partial_kz_1 + z_2^*\partial_iz_1^*\partial_jz_1\partial_kz_2 \nonumber\\
&= 2(z_1^*\partial_iz_1\partial_jz_2^*\partial_kz_2 + z_2^*\partial_iz_2\partial_jz_1^*\partial_kz_1)=2W_1,
\end{align}
where we again use the anti-symmetry of the levi-civita to swap the indices or the derivatives, e.g.\
$z_1^* \partial_iz_2^* \partial_jz_2 \partial_kz_1$ $= -z_1^* \partial_kz_2^* \partial_jz_2 \partial_iz_1
=z_1^* \partial_jz_2^* \partial_kz_2 \partial_iz_1$. We then prove that the contributions to the winding number (\ref{eq-Winding-4terms}) coming from the band mixing terms are twice the contribution resulting from a single eigenstate, $W_2=2W_1$, so that $W_1+W_2+W_3+W_4=6W_1$. With this, we conclude our proof that the winding number is equal to the Hopf invariant,
\begin{equation}\label{eq-W=H}
W= -\frac{1}{4\pi^2}
\int d^3p\; \varepsilon^{ijk}\; \Psi_1^\dag \partial_{i}\Psi_1 \partial_{j}\Psi_1^\dag \partial_{k}\Psi_1=H.
\end{equation}
Consequently, the periodically driven system can be topologically characterized in terms of a Hopf map where the Hopf invariant reveals the number of edge states in the quasienergy spectrum. Note that one can arrive at the same conclusion also through the approach followed by Schuster et al.\ in the supplementary of Ref.~[50], by writing the evolution operator as $U(\bk,t)=e^{i\phi(\bk,t)}|\Psi(\bk,t)\ra\la\Psi(\bk,t)| + e^{-i\phi(\bk,t)}(1-|\Psi(\bk,t)\ra\la\Psi(\bk,t)|)$ where $\pm\phi(\bk,t)$ are the phasebands for the two bands symmetric around zero due to the particle-hole symmetry.

\subsection{B. Experimental Scheme}
As is the case for a static system where the energy bands can be always made flat by tuning some parameters in the Hamiltonian adiabatically without closing any gaps, the quasienergy bands of a periodical drive can be flatten as well. This can be understood most easily in terms of the driving frequency. In Fig.S1, we consider the periodic drive depicted in Fig.1(c) of the manuscript and adiabatically connect it to the flat bands given in Fig.3(a). We plot the spectrum in a strip geometry to accentuate that the winding numbers remain same throughout the flattening, $W_{0,\pi}=1$, as visible with edge states present in each gap. As the driving frequency is lowered from $\omega=1.5J$ at the fixed sublattice offset $\Delta=0.1J$ in Fig.S1(a), the quasienergy bands become narrower. Note that, here the sublattice offset needs to vanish as well in order to obtain completely flat bands [Fig.S1(b)], illustrating other parameters that can be tuned in the system. In this example, the bands can be flatten within the entire BZ for a global frequency and offset, i.e.~$|\varepsilon(\mathbf{k})T|=\pi/2,\,\forall \mathbf{k}$ for period $T=3\pi/2J$. In general, it is possible to obtain $|\varepsilon(\mathbf{k})T(\mathbf{k})|=\pi/2$ by tuning the driving frequency individually at each $\mathbf{k}$, corresponding to a scaling of the period $T(\mathbf{k})$.

\begin{figure}[hb]
	\centering\includegraphics[width=.8\linewidth]{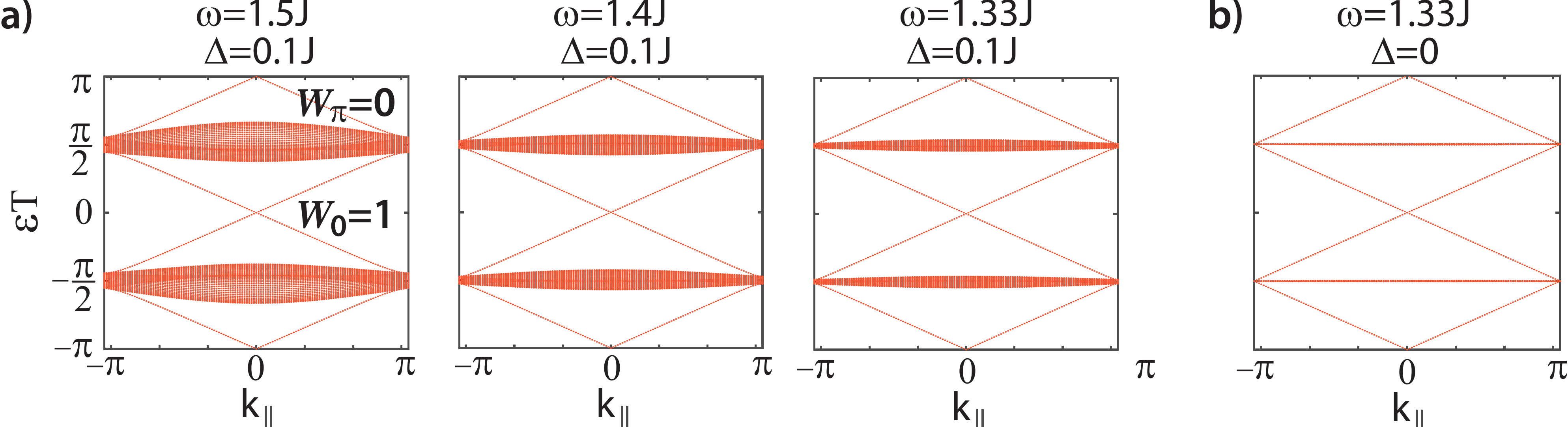}
	\caption{ Flattening of the quasienergy bands.}
\end{figure}

We now turn to folding of the Floquet spectrum, where it is instructive to consider the real-space picture. Let us imagine that we start with an initial state completely localized in one sublattice, say sublattice $A$. We then suddenly switch the drive on for the parameters $TJ/3=\pi/2,\,\Delta=0$, which correspond to the flat bands given in Fig.S1(b). During the first stage of the drive, particles are allowed to tunnel to the neighboring $B$ sites along the $\mathbf{b}_1$ direction (see the main text for the definition of the lattice directions). At the end of the first stage, particles complete a full Rabi population-transfer to $B$ sites for the time $TJ/3=\pi/2$. Similarly, during the second stage, the population transfers to $A$ sites along the $\mathbf{b}_2$ direction, then during the third stage to $B$ sites along the $\mathbf{b}_3$ direction. When the period is completed, particles end up completely localized on $B$ sites across the hexagon. Namely, this corresponds to flat bands which are symmetric around zero energy due to sublattice symmetry, but not degenerate since we started from $A$ sublattice and finished on $B$. However, if we keep monitoring the tunneling for one more period, particles keep circling around the hexagon and at $t=2T$, return to their starting point $A$. In other words, the time evolution operator over a double period becomes the identity, $\mathcal{U}(2T)=\mathds{1}$, corresponding to degenerate bands over the double period [Fig.3(b)] and giving access to the micromotion.

\end{document}